\documentclass[paper]{JHEP}
\usepackage{epsfig}
\usepackage{amssymb}
\def\as{\alpha_{\mbox{\tiny S}}}

\def\frac#1#2{ {{#1} \over {#2} }}

\def\beq{\begin{equation}}
\def\beqn{\begin{eqnarray}}
\def\eeq{\end{equation}}
\def\eeqn{\end{eqnarray}}

\newcommand\MCatNLO{{\rm MC}@{\rm NLO}}
\newcommand\SMCpNLO{{\rm SMC}$+${\rm NLO}}

\newcommand\pt{p_{\rm \scriptscriptstyle T}}
\newcommand\kt{k_{\rm \scriptscriptstyle T}}

\newcommand\bS{{\mathbb S}}
\newcommand\bI{{\langle \mathbb I|}}
\newcommand\Cf{C_{\rm F}}
\preprint{
 Bicocca--FT--04--11
}
\title{A New Method for Combining NLO QCD with Shower Monte Carlo Algorithms.}
\author{Paolo Nason\\
  INFN, Sezione di Milano,
  Piazza della Scienza 3, 20126 Milan, Italy\\
  E-mail: \email{Paolo.Nason@mib.infn.it}}
\abstract{
 I show that with simple extensions
 of the shower algorithms in Monte Carlo programs,
 one can implement NLO corrections to the hardest emission that
 overcome the problems of negative weighted events found
 in previous implementations.
 Simple variants of the same method can be used for an improved
 treatment of matrix element corrections in Shower Monte Carlo programs.}
\keywords{QCD, Monte Carlo, NLO Computations, Resummation,
Collider Physics}
\begin{document}
\section{Introduction}
Fixed order calculations in Perturbative QCD (PQCD) can be used
to compute inclusive quantities in strong processes.
The precision of these computations is limited by our ability
to compute complex Feynman graphs. Yet,
even if we could perform computations at arbitrary order, we would
not be able to give predictions for exclusive quantities.
This known fact follows from the presence of collinear and
soft divergences in fixed order calculations with a definite final state.
Only by summing over different final states these
divergences can cancel, thereby allowing the computation of certain
(i.e. the collinear and infrared insensitive) inclusive quantities.

The only frameworks in which exclusive quantities can be computed
must involve the resummation of an infinite class of Feynman graphs.
Shower Monte Carlo (SMC) programs perform this resummation in the
leading logarithmic (LL) approximation. Going beyond the LL
approximation is a considerably complex task:
the initial hard process would have to
be implemented at the NLO order, and shower development would have
to be improved to NLL accuracy, in both the collinear and soft
structure.
There are however simpler directions in which SMC
programs could be improved. Namely, one could stick to the LL
approximation as far as the shower development is concerned,
but improve the treatment of the hard emission,
in such a way that inclusive quantities are predicted with
a next-to-leading order (NLO) accuracy. Such an improvement
(I will call it {\SMCpNLO} from now on)
 would
clearly merge all the best features of the two approaches.
One would have exclusive final state generation together with
the accuracy of NLO calculations.

In refs.~\cite{Frixione:2002ik,Frixione:2003ei,Frixione:2003ep}
a method and implementation of an {\SMCpNLO}
program has been given.
Such method (referred to as \MCatNLO)
is based upon a careful elaboration of the NLO results,
that has to match certain features of the SMC program.
The approximate SMC implementation of the NLO corrections must
be subtracted to the exact NLO result in order to avoid double counting.
The method of refs.~\cite{Frixione:2002ik,Frixione:2003ei,Frixione:2003ep}
does not require modifications of the existing
shower Monte Carlo code. This is a
considerable advantage,
because Monte Carlo programs are large and
complex, and modifying them is a major task.
The method has however few drawbacks:
\begin{itemize}
\item
The approach is specific to a particular shower Monte Carlo
implementation, since the form of the NLO result has to be
adapted to it. Similarly, one must extract the NLO terms
already present in the Monte Carlo in order to subtract them.
Variations in the Monte Carlo implementation (like
the new showering variables proposed in
refs.~\cite{Gieseke:2003rz,Gieseke:2003hm})
would require computing again both the modified NLO expression
and the subtraction term.
\item
The Monte Carlo may not be fully accurate in the treatment
of the soft region. In this case the difference
between the exact NLO result and its Monte Carlo approximation
may have left over singularities that need special treatment.
\item
The correction coming from the difference between exact NLO
and its MC approximation may be negative. Thus, negative weighted
events may be generated.
\end{itemize}

In the present work, I propose an {\SMCpNLO} method that improves over
the method proposed in
refs.~\cite{Frixione:2002ik,Frixione:2003ei,Frixione:2003ep} on the
aspects listed above. This method cannot be directly implemented
using existing SMC programs, since it requires minor modifications to the
shower development. On the other hand, it is quite
simple, and it can also be used to perform matrix element corrections in a more
consistent way. It is not unlikely that it may be generalized
to implement a SMC$+$NNLO method, when NNLO calculations will become
widely available. It would be desirable that new SMC implementations
would eventually provide facilities to implement the method proposed here.

The strategy followed in the present approach is the following.
If the hardest (i.e. the largest $\pt$) emission was generated
first in the SMC, one would just have to correct the first emission at
NLO in order to get an {\SMCpNLO} implementation. Unfortunately,
in general this is not the case. A consistent treatment of
soft emissions is essential in SMC programs in order to obtain
the correct results for hadron multiplicity distributions.
Large angle soft radiation from bunches of collinear partons
interfere destructively, and the resulting coherent emission
is described in SMC programs as arising from the parent
parton of the collinear bunch. The corresponding
angular ordered shower algorithms
can therefore generate soft emissions earlier in the
shower~\cite{Marchesini:1984bm},
so that very often the hardest emission is not the first.
The basic result of this work is the construction of
a shower algorithm which is fully equivalent to the
coherent angular ordered shower, but has the hardest
emission generated first. This construction for timelike
showers is described in sec.~\ref{sec:HardestEmission}.
There I show that the hardest emission should be generated
with a modified Sudakov form factor, and subsequent emissions
should be generated according to the standard algorithm
modified by a $p_T$ veto, so that they cannot be harder
than the (first generated) hardest emission.
Large angle, soft coherent radiation
from the particles arising from the hardest splitting
must be added, which will result in the presence of
truncated showers associated to that pair of particles.

The NLO accurate hardest emission can be easily obtained,
as I show in Sec.~\ref{sec:MCatNLO}.
One simply constructs an NLO Sudakov form factor for the hardest
emission in analogy with the corresponding expression
in the SMC.

Space-like initial state showers are discussed
in detail in Sec.~\ref{sec:initshowers}.
They are usually implemented in the backward evolution
formalism \cite{Sjostrand:1985xi}, i.e. evolving from the
hard event back to the incoming hadron.
The shower order is from hard to
soft, as in the timelike case. This fact allows
a similar treatment of the spacelike and timelike cases.
Certain specific problems in the spacelike case, having to
do with the choice of the parton density scheme, are
discussed in detail and shown to be harmless.

The paper is organized as follows.  In
Sec.~\ref{sec:angularorderedshower} I collect the main formulae for
the angular ordered shower. In Sec.~\ref{sec:Shower} I
introduce a formal definition of the Monte Carlo shower, that will be
used in sec.~\ref{sec:HardestEmission} to construct the equivalent
formulation of the shower in which the largest $\pt$ emission is
generated first. In Sec.~\ref{sec:MCatNLO} the NLO implementation of
the hardest emission is discussed.  In Sec.~\ref{sec:initshowers} the
spacelike showers will be discussed.
 In Sec.~\ref{sec:other}, the
comparison with other matrix element corrections methods will be
discussed.  In Sec.~\ref{sec:conclusions} I give my conclusions and
prospects.

\section{The angular ordered shower}\label{sec:angularorderedshower}
\subsection{Kinematics}
\begin{figure}[htb]
\begin{center}
\epsfig{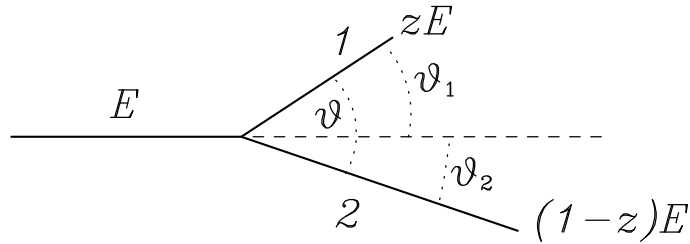}
\end{center}
\caption{\label{fig:splittingprocess}
Kinematic variables for a splitting process.}
\end{figure}

Shower Monte Carlo algorithms are an effective way of summing up a
large class of Feynman graphs in hard QCD collision, including
collinear singularities at the leading logarithmic level, and also
soft singularities in the double logarithmic region.  The
implementation given in the HERWIG Monte Carlo program is through an
angular ordered shower \cite{Marchesini:1984bm,Marchesini:1988cf}. We
summarize here its general features. We
consider the splitting process shown in
fig.~\ref{fig:splittingprocess}.
In the limit of small emission angles we have
\begin{equation}
p_T=\theta_1zE=\theta_2(1-z)E\,,\quad\quad \theta=\frac{\theta_1}{1-z}=\frac{\theta_2}{z}\,.
\end{equation}
Defining $t=E^2\theta^2$ we then have
\begin{equation}\label{eq:ptvalue}
p_T=\sqrt{t} z(1-z)\,.
\end{equation}
The probability for such a split is given by
\begin{equation}\label{eq:splittingker}
dP=F(z,t) dz\,dt=\frac{\alpha_S(p_T)}{2\pi}\hat{P}_{ij}(z)\frac{dt}{t} dz
\end{equation}
where $\hat{P}$ is the unregularized Altarelli-Parisi splitting function.
The variable $z$ is limited by the requirement $p_T=\sqrt{t}\, z(1-z) >
\sqrt{t_0}/4$, and $t_0$ is the minimum $t$ for a branching to take
place. In the angular ordered shower, subsequent branching can take
place only with an angle $\theta$ less than the previous branching
angle.  In terms of $t$, this implies that the maximum $t$ value for a
subsequent splitting in fig.~\ref{fig:splittingprocess} must be less
than $z^2t$ along line 1, and $(1-z)^2 t$ along line 2.
\subsection{Coherence}
\begin{figure}[htb]
\begin{center}
\epsfig{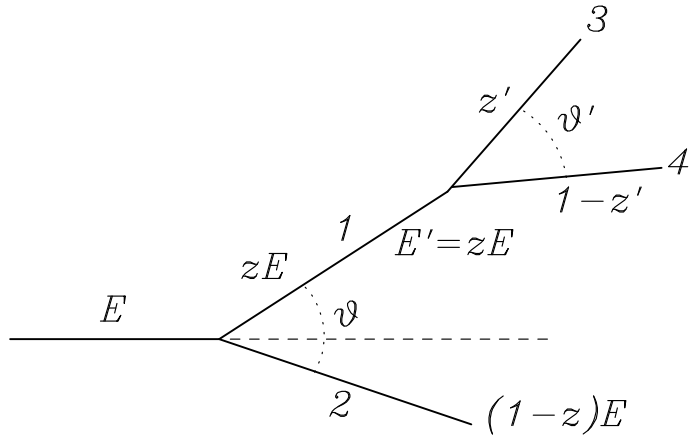}
\end{center}
\caption{\label{fig:twobranchings}
Two subsequent branchings.}
\end{figure}

In the angular ordered shower it is quite possible to have a large
$p_T$ emission following smaller $p_T$ emissions. In the sequence of
two branchings depicted in fig.~\ref{fig:twobranchings}, for example,
one can have $z$ very near one in the first branching, and $z^\prime=1/2$ in
the second branching, so that there is a region of
$\theta^\prime<\theta$ for which the second $p_T$ (according to formula
(\ref{eq:ptvalue})) is much larger than the first one.
The line $(1-z)E$ would be soft in this case, and (if line 2 is a gluon)
the splitting kernel eq.~(\ref{eq:splittingker}) is singular in this region.
Thus it would appear that a highly off-shell line can
have soft gluon emission singularity. In fact
this is not what happens.
The soft gluon emission in this region
is the result of coherent radiation from lines 3 and 4,
and these line have in fact small virtualities.

\section{The shower}\label{sec:Shower}
\subsection{Notation}
We now want to set up some notation for discussing the properties of a
shower.  We call $\bS(t)$ the angular ordered shower originating from the
initial value $t$. In order to give a more precise meaning to this notion
we introduce the basic configuration states\footnote{We stress that these
states do not represent quantum states.}
\begin{equation}
|k_1,m_1; ... ;k_l,m_l\rangle
\end{equation}
where $k_i$ and $m_i$ are the momenta and quantum numbers
of the particles, with the normalization\footnote{We do not need
to symmetrize the states over identical particles. Physical
observables will always be symmetric.}
\begin{equation}
\langle k_1,m_1; ... ;k_l,m_l|k_1^\prime,m_1^\prime;
 ... ;k_{l^\prime}^\prime,m_{l^\prime}^\prime\rangle =
\delta_{l,l^\prime}
 \prod_{i=1}^l \delta^3(k_i-k^\prime_i)
\delta_{m_i,m^\prime_i}\;.
\end{equation}
A shower is defined as
\begin{equation}
\bS = \sum_{l=1}^\infty \sum_{m_1 \ldots m_l}
\int d^3 k_1\ldots d^3 k_l \;
 C(k_1,m_1; \ldots ; k_l,m_l) \;
\langle k_1,m_1; ... ;k_l,m_l|
\end{equation}
so that given the infinitesimal cell around a basic state
\begin{equation}
d\Psi=|k_1^\prime,m_1^\prime;  ... ;k_{l^\prime}^\prime,m_{l^\prime}^\prime\rangle \;
 d^3 k_1^\prime \ldots d^3 k_{l^\prime}^\prime
\end{equation}
the product
\begin{equation}
S \cdot d\Psi
\end{equation}
is the probability to generate a state in the cell $d\Psi$.
A final state observable $g(k_1,m_1; \ldots ; k_l,m_l)$ will
be represented in this notation by
\begin{equation}
{\mathbb G}=  \sum_{l=1}^\infty \sum_{m_1 \ldots m_l}
\int d^3 k_1\ldots d^3 k_l \;
 g(k_1,m_1; \ldots ; k_l,m_l)\;
|k_1,m_1; k_2,m_2; ... ;k_l,m_l\rangle
\end{equation}
and the shower is a linear functional over the observable, which
yields its average value $\bS \cdot {\mathbb G}$.
\subsection{The shower equation}
The angular ordered shower obeys the equation
\begin{equation}\label{eq:herwigequation}
\bS(t_I)=\Delta(t_I,t_0)\,\bI + \int_{t_0}^{t_I} \Delta(t_I,t)\,F(z,t)\,
\bS(z^2 t)\, \bS(\,(1-z)^2\, t\,)\, dt\,dz
\end{equation}
where $\bI$ stands for the initial parton $\langle k,m|$,
and $\Delta(t_1,t_2)$
is the probability that no emission takes
place with the ordering variable between $t_1$ and $t_2$.
The first term represents the ``no emission'' case.
When we write the product of two showers
$\bS(z^2 t)\, \bS(\,(1-z)^2\, t\,)$
we just mean the obvious thing, i.e. the formal product where
the product of basic states is interpreted as a tensor product.
We observe that eq.~(\ref{eq:herwigequation}) is fully meaningful
only if, when solved by iteration, it is possible to reconstruct its
final state. Thus, we implicitly assume that the parton energy and
direction is specified in the shower, and furthermore, that the azimuthal
angle in the splitting also appears in the integration.
We do not write explicitly these dependences in order to have a lighter
notation.

Equation (\ref{eq:herwigequation}) can be written graphically as
\begin{equation}\label{eq:graphicherwigequation}
\mbox{\epsfig{file=showereq.eps,width=0.5\textwidth}}
\end{equation}
where the large blobs stands for the shower, thick lines are the Sudakov
form factors, and the small blob is the splitting function. The dashed
lines represent the connections of the shower blobs to the vertices, and
no factors are attached to them. We have omitted to write indices
along the lines and vertices of the shower, specifying particle
species. They are inessential in the following arguments, and therefore
we carry out our discussion assuming that we re dealing with gluons
only.
\subsection{Unitarity}
Shower Monte Carlos guarantee that given the initial $t_I$ one
and only one configuration is reached.
Thus the observable ${\mathbb G}$ with $g(k_1,m_1; \ldots ; k_l,m_l)=1$
for all $l$, $k$, and $m$ must yield $\bS\cdot {\mathbb G}=1$.
In other words, unitarity must be satisfied:
the sum of the probabilities of all shower configurations
must yield one.
Thus, from eq.~(\ref{eq:herwigequation}) we immediately obtain
\begin{equation}\label{eq:herwigunitarity}
1=\Delta(t_I,t_0) + \int_{t_0}^{t_I} \Delta(t_I,t)\,F(z,t)\, dt\,dz\;,
\end{equation}
and with the ansatz
\begin{equation}
\Delta(t_1,t_2)=\frac{\Delta(t_1,t_0)}{\Delta(t_2,t_0)}\,,
\end{equation}
we get the solution
\begin{equation}
\Delta(t,t_0)=e^{-\int_{t_0}^t F(z,t^\prime) \,dz\,dt^\prime},
\end{equation}
which is the usual expression for the Sudakov form factor.
Notice that unitarity works only if
the expression $\Delta(t_I,t)\,F(z,t) dt dz$
is an exact differential. This is
equivalent to saying that $\Delta(t_I,t)$ is the probability that
no emission has taken place between $t_I$ and $t$. Under this condition,
the usual probabilistic algorithm of shower Monte Carlo can
be used to generate events, i.e. one picks a random number $r$ between
0 and 1, solves the equation $\Delta(t_I,t)=r$ for $t$, and then
generates $z$ with a distribution proportional to $F(z,t)$.
\subsection{Vetoed showers}
In order to become familiar with the notation we will now prove that
the standard vetoing procedure\footnote{See for example ref.~\cite{Seymour:1994df}.}
is equivalent to the use of a modified
Sudakov form factor. We will use this well known result in the
following sections, and thus include its proof for
completeness.

 A vetoed shower is obtained by applying a
constraint $\theta(g(z,t))$ to the branching process. One generates a
branching at a scale $t$, and, if $g(z,t)$ is negative, the
branching is rejected, and one generates a new branching starting from
$t$.  The equation for a vetoed shower is thus
\begin{eqnarray}
\bS_V(t_I)&=&\Delta(t_I,t_0)\,\bI + \int_{t_0}^{t_I} \Delta(t_I,t)\,F(z,t)\,
\theta(g(z,t))\,
\bS_V(z^2 t)\, \bS_V(\,(1-z)^2\, t\,)\, dt\,dz
\nonumber \\\label{eq:vetoedherwigequation}
&+&
\int_{t_0}^{t_I} \Delta(t_I,t)\,F(z,t)\,
\left[1-\theta(g(z,t))\right]\, \bS_V(t)\, dt\,dz\;.
\end{eqnarray}
Using the identity
\begin{equation}
\Delta(t_I,t)=\frac{\Delta(t_I,t_0)}{\Delta(t,t_0)}\;,
\end{equation}
dividing both sides by $\Delta(t_I,t_0)$ and taking a derivative
with respect to $t_I$ we get
\begin{eqnarray}
\frac{\partial}{\partial t_I} \frac{\bS_V(t_I)}{\Delta(t_I,t_0)}&=&\frac{1}{\Delta(t_I,t_0)}
\int F(z,t_I)\,\theta(g(z,t_I))\,
\bS_V(z^2 t_I)\, \bS_V(\,(1-z)^2\, t_I\,)\, dz
\nonumber \\\label{eq:vetoedherwigequation1}
&+&
\frac{\bS_V(t_I)}{\Delta(t_I,t_0)}\int F(z,t_I)\,
\left[1-\theta(g(z,t_I))\right]\, dz\;.
\end{eqnarray}
Defining now
\begin{equation}
r(t_I,t_0)=\exp\left(-\int_{t_0}^{t_I} F(z,t)\,
\left[1-\theta(g(z,t))\right]\,dt\, dz\right)\;,
\end{equation}
we have
\begin{equation}
\frac{\partial r(t_I,t_0)}{\partial t_I}=-r(t_I,t_0) \int F(z,t_I)\,
\left[1-\theta(g(z,t_I))\right]\,dz\;,
\end{equation}
so that eq.~(\ref{eq:vetoedherwigequation1}) becomes equivalent to
\begin{equation}\label{eq:vetoedherwigequation2}
\frac{\partial}{\partial t_I} \frac{r(t_I,t_0)\,\bS_V(t_I)}{\Delta(t_I,t_0)}=
\frac{r(t_I,t_0)}{\Delta(t_I,t_0)}\int F(z,t_I)\,
\theta(g(z,t_I))\,
\bS_V(z^2 t_I)\, \bS_V(\,(1-z)^2\, t_I\,)\, dz\;.
\end{equation}
Defining now
\begin{equation}
\Delta^\prime(t_I,t_0)=\frac{\Delta(t_I,t_0)}{r(t_I,t_0)}=
\exp\left( -\int_{t_0}^{t_I} F(z,t)\,
\theta(g(z,t))\,dt\, dz\right)\;,
\end{equation}
and integrating eq.~(\ref{eq:vetoedherwigequation2}) using the initial
condition of eq.~(\ref{eq:vetoedherwigequation})
we get
\begin{equation}
\bS_V(t_I)=\Delta^\prime(t_I,t_0)\,\bI + \int_{t_0}^{t_I} \Delta^\prime(t_I,t)\,F(z,t)\,
\theta(g(z,t))\,
\bS_V(z^2 t)\, \bS_V(\,(1-z)^2\, t\,)\, dt\,dz\;.
\end{equation}
which is a shower equation with a modified Sudakov form factor which
matches precisely the splitting vertex.

In the following, we will encounter showers with Sudakov form factors
and splitting vertices containing a theta function that limits the
transverse momentum.  These showers can
be simply implemented by vetoing, without the need to
modify the form factor calculation procedure in a SMC.
\section{The largest $\pt$ emission}\label{sec:HardestEmission}
\begin{figure}[htb]
\begin{center}
\epsfig{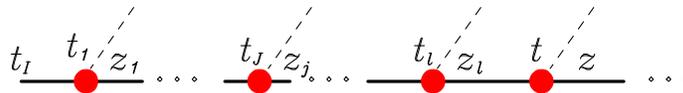}
\end{center}
\caption{\label{fig:largestptseq}
Hardest $\pt$ occurring at $t,z$, in a shower initiated at $t_I$.}
\end{figure}

We would like now to transform the angular ordered equation
in such a way that the hardest emission is generated first.
The first step to perform such transformation is to prove the
following three statements:
\begin{itemize}
\item[(I)] \emph{the largest $\pt$ emission in an angular
ordered shower always takes place along the hardest line in the
shower}. (The hardest line in the shower is the one that starts at
the beginning of the shower and always follows the largest $z$
line in the splitting vertices.)
\item[(II)] \emph{along the hardest line, configurations with non-soft emission
before the hardest emission are collinear subleading}.
\item[(III)] \emph{along the hardest line, for leading configurations,
from the first non-soft emission down to the last we have $t\lesssim \pt^2$
(i.e. $t$ of order $\pt^2$ or smaller)}.
\end{itemize}
The proof of (I) is quite simple.
Let us call $t,z$ the emission variables of the largest $\pt$ splitting.
There is a unique way to walk backward in the shower, up to the initial
$t_I$. Let us call $t_1,z_1$, \ldots $t_l,z_l$ all the splitting
occurring along this line before $t,z$,
according to fig.~\ref{fig:largestptseq}.
Consider any splitting $t_j,z_j$ before $t,z$.
We must have
$t <z_j^2 t_j$. If $\pt>{\pt}_j$ then
$t z^2(1-z)^2>t_j z_j^2(1-z_j)^2$, which implies
$(t_jz_j^2) z^2(1-z)^2>t_j z_j^2(1-z_j)^2$, and thus
$z^2(1-z)^2>(1-z_j)^2$, which immediately implies
$1-z_j<1/4$, and thus $z_j>3/4$ for all $j$, which proves our lemma.

The proof of (II) goes as follows. Suppose there is a non-soft
emission at $t_k,z_k$ before the hardest emission. Non-soft means $1-z_k\approx 1$.
Also $z_k>1/2$, therefore from $\sqrt{t_k}z_k(1-z_k)<\pt$ we get $t_k\lesssim \pt^2$
and \emph{a fortiori} $t_k \lesssim t$. But by angular ordering we must have $t_k>t$.
Thus none of the collinear integrals from $t_k$ up to $t$ has a large logarithmic
range, and so these configurations are subleading.

The proof of (III) goes as follows. The first non-soft emission must be the hardest
emission itself, or may occur after it, but not earlier because of statement (II).
Let us call $t^\prime,z^\prime$ the values corresponding to the first non-soft
emission. We must have $\sqrt{t^\prime} z^\prime(1-z^\prime)<\pt$,
which implies $t^\prime \lesssim \pt^2$. By angular ordering the same must hold for all
following emissions.

The equation for the angular ordered shower (\ref{eq:herwigequation})
can be formally solved by iteration. Using the fact that the largest
$\pt$ emission has to be on the hardest line, we can write a formal
solution where the largest $\pt$ is explicitly present:
\begin{equation}\label{eq:largestptexp}
\bS(t_I)=\Delta(t_I,t_0)\bI+
\sum_{l,k=0}^{\infty}\int
\mbox{\epsfig{file=largestptexp.eps,,width=0.6\textwidth}}.
\end{equation}
The $z,t$ vertex has the largest $\pt$.

The graphic expression in fig.~\ref{eq:largestptexp} has the following
meaning\footnote{ In the following we will also
denote collectively the whole $z,t$ variable sequence as $z_i,t_i$
with $i=1,\ldots,l+k+1$, where $z,t=z_{l+1},t_{l+1}$, and
$z_{l+1+j},t_{l+1+j}=\tilde{z}_j,\tilde{t}_j$.}:
\begin{itemize}
\item
Thick lines are Sudakov form factors; more specifically,
they are given by
\begin{equation}
 \Delta(t_I,t_1),\,\Delta(z_1^2 t_1,t_2),\ldots,
\Delta(z_{l+k+1}^2 t_{l+k+1},t_0)\,.
\end{equation}
\item
The large solid blobs are $\bS((1-z_i)^2 t_i^2)$,
\item
The hollow vertex blobs stand for
\begin{equation}
2F(z_i,t_i)\,
\theta(z_i-1/2)\,
\theta(\pt - \sqrt{t_i} z_i(1-z_i))\;.
\end{equation}
where we have defined $\pt=\sqrt{t}z(1-z)$.
\item
The solid vertex blob (i.e. the largest $\pt$ vertex) is given by the same
expression as the hollow blob without the theta function.
\item
All intermediate $z_i,t_i$ are integrated.
\end{itemize}
The factor of 2 accounts for the fact that the largest $\pt$ can occur
on either line of the splitting.
Larger $\pt$ emissions cannot occur along the $1-z_i$ (and $1-\tilde{z}_i$)
lines, according to our statement (I).

Equation (\ref{eq:largestptexp}) cannot be implemented as it stands in a Monte Carlo
algorithm: the Sudakov form factors do not match with the splitting vertices,
because of the $\pt$ theta function. We thus perform the following
manipulation. We rewrite the Sudakov form factors as
\begin{eqnarray}\label{eq:sudsplit}
&&\Delta(t_iz_i^2,t_{i+1})=
e^{-\int_{t_{i+1}}^{t_iz_i^2} dt^\prime\int dz^\prime
 F(t^\prime,z^\prime)}
 \\
&=& e^{-\int_{t_{i+1}}^{t_iz_i^2} dt^\prime\int_{1/2}^1
 dz^\prime
2F(t^\prime,z^\prime)\,\theta(\pt-z^\prime(1-z^\prime)\sqrt{t^\prime})}
 \times
e^{-\int_{t_{i+1}}^{t_iz_i^2} dt^\prime\int_{1/2}^1
dz^\prime
2 F(t^\prime,z^\prime) \theta(z^\prime(1-z^\prime)\sqrt{t^\prime}-\pt)}\;.
\nonumber
\end{eqnarray}
The first factor on the right hand side of eq.~(\ref{eq:sudsplit})
matches precisely the splitting vertex in eq.~(\ref{eq:largestptexp}).
The second factor is a remnant.
Because of statements (II) and (III) the remnant can always be written as
\begin{equation} \label{eq:remnantform}
e^{-\int_{t_{i+1}}^{t_iz_i^2} dt^\prime\int_{1/2}^1
dz^\prime
2 F(t^\prime,z^\prime) \theta(z^\prime(1-z^\prime)\sqrt{t^\prime}-\pt)}
\approx 
e^{-\int_{t_{i+1}}^{t_i} dt^\prime\int
dz^\prime
F(t^\prime,z^\prime) \theta(z^\prime(1-z^\prime)\sqrt{t^\prime}-\pt)}\;.
\end{equation}
In fact, assume that $k$ is the first non-soft vertex. Because of statement (II)
in all previous vertices we can replace $z_i\to 1$, and eq.~(\ref{eq:remnantform}) holds.
According to statement (III), starting with
the first non-soft vertex the $t$ values are not larger than $\pt$, so that the
exponent in eq.~(\ref{eq:remnantform}) vanishes because of the theta function,
and also in this case eq.~(\ref{eq:remnantform}) is trivially satisfied.

We will call $\Delta_R(t_I,\pt)$ the product of all remnants.
From eq.~(\ref{eq:remnantform}) we obtain
\begin{equation} \label{eq:deltaR}
\Delta_R(t_I,\pt) \equiv 
=e^{-\int_{t_0}^{t_I} dt^\prime\int
dz^\prime
F(t^\prime,z^\prime) \theta(z^\prime(1-z^\prime)\sqrt{t^\prime}-\pt)}\;.
\end{equation}
Equation (\ref{eq:deltaR}) has an obvious physical interpretation.
It is the Sudakov form factor for not emitting a particle with
transverse momentum larger than $\pt$.
It can be used to generate
the largest $\pt$ emission first, in association with the
$F(z,t)\, dz\, dt$ factor present in eq.~(\ref{eq:largestptexp}).
We can thus rewrite eq.~(\ref{eq:largestptexp}) in the following form
\begin{eqnarray}\label{eq:largestptexpf}
&& \bS(t_I)=\Delta(t_I,t_0)\bI+
\int_{t_0}^{t_I} dt \,\theta(z>1/2)\, dz\,
\Delta_R(t_I,\pt) 2F(z,t)\;\bS((1-z)^2t)
\nonumber \\
&&
\times
\sum_{l=0}^\infty
\mbox{\epsfig{file=softsh.eps,height=2cm}}\;
\times
\sum_{k=0}^{\infty}
\mbox{\epsfig{file=finsh.eps,height=2cm}}\;.
\end{eqnarray}
Double lines are the modified Sudakov form factors,
defined as
\begin{equation}
\Delta_V(t_i z_i^2,t_{i+1})=
 e^{-\int_{t_{i+1}}^{t_iz_i^2} dt^\prime\int_{1/2}^1
 dz^\prime
2F(t^\prime,z^\prime)\,\theta(\pt-z^\prime(1-z^\prime)\sqrt{t^\prime})}\;.
\end{equation}
The two graphic factors in eq.~(\ref{eq:largestptexpf}) are now
showers, that can be easily implemented in a Monte Carlo algorithm.
In fact they are angular ordered showers with the addition of a
$\pt$ veto. The veto needs to be applied only along the hardest line,
but since no emissions with transverse momenta larger than $\pt$ are
possible in any other branches of the showers, we might as well apply
the veto everywhere, without the need to follow the hardest line.

Our final result for the shower is
\begin{eqnarray}\label{eq:largestptexpfin}
&& \bS(t_I)=\Delta(t_I,t_0)\bI +
\int_{t_0}^{t_I}dt\int_{0}^1 dz \, \Delta_R(t_I,\pt) F(z,t)\;\bS_V((1-z)^2t,\pt)
\nonumber \\
&&
\times
\bS_{VT}(t_I,t,\pt)\;\bS_V(z^2 t,\pt)\;,
\end{eqnarray}
where
\begin{itemize}
\item $\bS_V((1-z)^2 t,\pt)$ and $\bS_V(z^2 t,\pt)$ are angular ordered
$\pt$ vetoed showers.
\item $\bS_{VT}(t_I,t,\pt)$ is also an angular ordered
$\pt$ vetoed shower, starting at $t_I$,
except that along the hardest line the minimal scale for showering
is $t$ instead of $t_0$. We call it the truncated shower.
\end{itemize}
Unitarity of eq.~(\ref{eq:largestptexpfin})
works as follows.
Since both $\bS_{VT}(t_I,t,\pt)$
and $\bS_V(t,t_0)$ project to 1 by unitarity, the unitarity equation is
\begin{equation}
1=\Delta(t_I,t_0) +
\int_{t_0}^{t_I}dt\int_{0}^1 dz \, \Delta_R(t_I,\pt) F(z,t)\;,
\end{equation}
which is the analogue of eq.~(\ref{eq:herwigunitarity}).
We have
\begin{eqnarray}
&&\int_{t_0}^{t_I}dt\int_{0}^1 dz \, \Delta_R(t_I,\pt) F(z,t)
\nonumber \\
&&=\int d\pt \int_{t_0}^{t_I}dt\int_{0}^1 dz\, 
\delta(\pt-\sqrt{t}z(1-z))\, F(z,t)
e^{-\int_{t_0}^{t_I} dt^\prime\int
dz^\prime
F(t^\prime,z^\prime) \theta(z^\prime(1-z^\prime)\sqrt{t^\prime}-\pt)}
\nonumber \\
&&=\int d\pt \frac{d}{d\pt} e^{-\int_{t_0}^{t_I} dt^\prime\int
dz^\prime
F(t^\prime,z^\prime) \theta(z^\prime(1-z^\prime)\sqrt{t^\prime}-\pt)}
=1-\Delta(t_I,t_0)\;.
\end{eqnarray}
The last step follows from the fact that the value of the exponential
is 1 for $\pt\to\infty$ (because the exponent vanishes),
and it reduces to $\Delta(t_I,t_0)$
for $\pt=0$, since in this case the theta function is irrelevant.

The first emission in eq.~(\ref{eq:largestptexpfin}) is given in terms
of a Sudakov form factor that depends upon $z$ and $t$, through the
$\pt$ variable. It can be easily expressed in terms of $\pt$ as
an independent variable
\begin{equation}
\int_{t_0}^{t_I}\frac{dt}{t}\int_{0}^1 dz \Longrightarrow
\int_{\sqrt{t_0}/4}^{\sqrt{t_I}/4}\frac{2\, d\pt}{\pt} \int_{0}^1 dz
\;\theta(\sqrt{t_I}-t)\;,
\end{equation}
where now $t=\pt^2\,/\,z^2(1-z)^2$ is the dependent variable.
Observe that the lower bound in $\pt$ is $\sqrt{t_0}/4$. As discussed earlier,
we always have an implicit cut-off at this value of $\pt$ that prevents
the argument of $\as$ from becoming too small.

We can express the first emission Sudakov form factor in
terms of a $\pt$ integral
\begin{equation}
\Delta_R(t_I,\pt)=\exp\left[-\int_{\pt}^{\sqrt{t_I}/4} d{\pt^{\prime}}
\frac{2t^\prime}{\pt^{\prime}} F(t^\prime,z^\prime)\theta(t_I-t^\prime)\right] \;.
\end{equation}
Now $\Delta_R(t_I,\pt)$ can be used to generate the largest $\pt$
emission using the standard SMC technique: one generates a random number
$0<r<1$ and solves the equation $\Delta_R(t_I,\pt)$ for $\pt$.
Alternatively, one can define
\begin{equation}
\tilde{\Delta}_R(\pt^{\rm max},\pt)=\exp\left[-\int_{\pt}^{\pt^{\rm max}}
 d{\pt^{\prime}}
\frac{2t^\prime}{\pt^{\prime}} F(t^\prime,z^\prime)\right] \;,
\end{equation}
generate the event according to $\tilde{\Delta}_R$, and implement the
$\theta(t_I-t^\prime)$ by vetoing.

The Sudakov form factors in $\bS_{VT}(t_I,t,\pt)$
and $\bS_V(t,\pt)$ include the $\pt$ theta function. They
depend therefore upon two variables, $t$ and $\pt$.
They can be
easily implemented using the standard Sudakov form factor (i.e. without
the theta function), and
a $\pt$ veto procedure as in eq.~(\ref{eq:vetoedherwigequation})
with $g(z,t)=\pt-\sqrt{t}z(1-z)$.

The truncated shower is a soft shower. It does not
appreciably degrade the energy entering the hardest emission.
This fact is necessary in order to generate the hardest emission
first. It is also natural: the radiation of the
truncated shower is in reality coherent radiation from
final state particles, and so it should not steal energy from the
hardest splitting.

Observe that hard radiation in the truncated shower does not have
any logarithmic enhancement (i.e. neither soft nor collinear).
 Formally, this works as follows. If $z$ is not near 1
we can only have collinear logarithms.
Because of the angular ordering and the $p_T$ veto we have
\begin{equation}
\theta<\theta_i\,,\quad E\,\theta_i z_i (1-z_i)
 < (z_i E)\,\theta\, z(1-z) <\frac{z_i\, E\theta}{4}\;
 \Longrightarrow\;
\theta<\theta_i<\frac{\theta}{4(1-z_i)}
\end{equation}
(which forces $z>3/4$). In order to have a logarithmic integral
we need a large $\theta_i$ range, and this is possible only if
$z\to 1$. Thus no logarithmic enhancement is present for
hard radiation.

\section{NLO Corrections}\label{sec:MCatNLO}
\subsection{NLO expansion of the SMC}
The argument given in the previous Section
was referring to a single jet being produced.
The extension to the general case is straightforward.
If there are $m$ primary partons, one introduces
the total Sudakov form factor
\begin{equation} \label{eq:multisud}
\Delta_R(\{t_I\},\pt)=\prod_{i=1,m} \Delta_R^i(t_I^i,\pt)\;,
\end{equation}
where with $\{t_I\}$ we denote the set of the initial showering
variables $t_I^i$, $i=1\ldots m$ for all primary partons.
The probability distributions for the hardest emission
is given by
\begin{eqnarray}
d\sigma &=& 
B(p_1 \ldots p_m) d\Phi_m\nonumber \\ \label{eq:hardestapprox}
&\times&\left[\Delta_R(\{t_I\},0)
+\Delta_R(\{t_I\},\pt) \sum_{i=1,m} F_i(z,t)\, \theta(t_I^i-t)\,dz\, dt
\,\frac{d\phi}{2\pi} \right]\;.
\end{eqnarray}
The derivation of equation (\ref{eq:hardestapprox}) is a straightforward
extension of the argument presented in Sec.~\ref{sec:HardestEmission}.
Notice that in each emission term in eq.~(\ref{eq:hardestapprox}),
according to eq.~(\ref{eq:multisud}),
there are $\Delta_R(t_i,\pt)$ factors also for the
primary partons that are not emitting. These factors compensates for the $\pt$
veto that is applied also to their showers.
 
The Monte Carlo generation of the event proceeds as follows.
One generates the hardest $\pt$ according to $\Delta_R(\{t_I\},\pt)$.
The $i$ and $z$ values\footnote{We assume for simplicity an isotropic
distribution in the azimuth $\phi$ of the radiated particle.}
are chosen with a probability proportional to $F_i(z,t)\, \theta(t_I^i-t)$,
where $t=\pt^2/z^2/(1-z)^2$.
Once $i$ and $z$ are chosen, we construct the two partons
originating from $i$. We let all the $m+1$ partons shower with a $\pt$
veto, and add a truncated shower, starting from $t_I^i$ down to $t$,
originating from $i$.

Equation~(\ref{eq:hardestapprox}) has the following ${\cal O}(\as)$
expansion
\begin{eqnarray}
d\sigma &=&
B(p_1 \ldots p_m) d\Phi_m \left[1-\sum_{i=1,m}\int_{t_0}^{t_I^i} F_i(z,t)\, dz\, dt
+\sum_{i=1,m} F_i(z,t)\, \theta(t_I^i-t)\, dz\,
\frac{d\phi}{2\pi} dt\right]
\nonumber \\\label{eq:mcnlo}
&=& B(p_1 \ldots p_m) d\Phi_m \left[1+
\sum_{i=1,m} F_i(z,t)_+\, \theta(t_I^i-t)\, dz\,\frac{d\phi}{2\pi} dt\right]
\end{eqnarray}
where the first two terms in the square bracket arise from the
expansion of $\Delta_R$.
The $+$ notation means that the singularities
in $t$ and $z$ are regulated according to the $+$ prescription, i.e.
in such a way that
\begin{equation}
\int_{t_0}^{t_I^i} dt dz  F_i(z,t)_+=0\;.
\end{equation}
The $+$ prescription guarantees that unitarity is preserved also at the perturbative
level, through the familiar cancellation of real and virtual singularities.
In fact, the second term in the square bracket of eq.~(\ref{eq:mcnlo})
is the virtual correction.
Observe that also in eq.~(\ref{eq:hardestapprox}),
by unitarity, the factor in the square bracket integrates to one, so that
the cancellation mechanism works to all order in the perturbative expansion.

\subsection{Parton event generator}
In a ``parton'' event generator the ${\cal O}(\as)$ approximation
to the radiation process would be implemented
by associating to an event with radiation, characterized
by the variables $p_1\ldots p_n,z,t,\phi$, a counter-event
with opposite (negative) weight, and $n$ body kinematics
$p_1\ldots p_n$. This association (or \emph{projection}) is
the inverse operation of what the SMC does when splitting one particle
into two. In the splitting process the SMC
substitutes the $i^{\rm th}$ particle in the ensemble
$p_1\ldots p_m$ with two new particles, with momenta computed as a function
of $p_i,z,t$ and $\phi$. After this, some momentum reshuffling is needed,
since the sum of the four-momenta of the splitting products now has
a non vanishing invariant mass, and energy-momentum conservation must be
enforced.

The SMC may generate radiation from different primary partons in
overlapping regions.  In this case a radiated parton may be associated
with more than one of the $m$ primary partons.  Thus, in general, for
a given event with one emission, there can be $m$ projections and $m$
counterterms, one for each association of the radiated parton with one
primary parton.

NLO calculations performed using the subtraction method often
associate to the event with radiation few counter-events with no
radiation, according to some projections that are chosen for
convenience\footnote{A specific illustration of this method is given
in ref.~\cite{Mangano:1991jk}, section 4.}.

The ${\cal O}(\as)$ expansion of the SMC given in term
of the subtraction method, as presented here, was first obtained in
ref.~\cite{Frixione:2002ik}, where it plays a fundamental role in
the computation of the term to subtract from the NLO cross section in
order to avoid overcounting. In the present work, our aim will be to
substitute the SMC NLO result with the exact one, and then
work our way backward to the (NLO accurate) analogue of
eq.~\ref{eq:hardestapprox}.
\subsection{Exact NLO formula}
The exact next-to-leading expression for the cross section
can be described schematically by the following formula
\begin{eqnarray}\label{eq:exactnlo}
d\sigma &=& B(p_1 \ldots p_m) d\Phi_m + V(p_1 \ldots p_m) d\Phi_m
\nonumber \\
&+&
\left[R(p_1 \ldots p_{m+1}) d\Phi_{m+1}
-\sum_i C_i(p_1 \ldots p_{m+1}) d\Phi_{m+1} {\mathbb P}_i
\right]
\end{eqnarray}
which should be read as follows: the Born and virtual terms generate
$m$ body events, the real emission term generates $m+1$ body events,
weighted by $R$,
with some associated counter-events with weights $C_i$, and $m$ body kinematics
obtained applying the projections ${\mathbb P}_i$ to the $m+1$ body event.
The projection is required to be IR and collinear
insensitive, so that if one parton becomes soft the projected
configuration is obtained from the full configuration by removing the
soft parton, and if two partons become collinear the projected
configuration is obtained by merging the two collinear partons.
Furthermore, the difference between $R$ and $\sum C_i$ is non-singular.
\subsection{{\SMCpNLO} strategies}
In the {\MCatNLO} implementation
\cite{Frixione:2002ik,Frixione:2003ei,Frixione:2003ep}, the
Monte Carlo hardest emission is corrected to match the NLO calculation.
This is done by the following procedure:
\begin{itemize}
\item[(i)]
The NLO cross section formula is rewritten using a projection $\mathbb P$
that coincides with the shower Monte Carlo projection.
\item[(ii)]
The shower Monte Carlo approximation to the NLO cross section,
eq.~(\ref{eq:mcnlo})
is subtracted to the exact NLO cross section (\ref{eq:exactnlo}).
This subtraction is possible
since the two formulae, after step (i), have the same projection.
\item[(iii)]
The Monte Carlo is run with its standard input, plus a correction
obtained by adding a contribution where the subtracted NLO correction
of step (ii) is used as an input to the Monte Carlo
shower.
\end{itemize}
The advantage of this method is that it does not require to modify the
showering code in order to be implemented. Observe that the NLO correction
in (ii) is always a hard correction, since collinear and soft singularities
have been removed from the NLO formula by subtracting the Monte Carlo
NLO approximation. The disadvantages of this approach are
\begin{itemize}
\item
The NLO calculation has to be tuned to the Monte Carlo, by changing
the projection $\mathbb P$. It becomes therefore Monte Carlo
dependent.
\item
One has to determine the Monte Carlo NLO
approximation, a task that is not always simple.
\item
The Monte Carlo implementation of the soft limit
may be less than perfect. Thus, one has to make sure that the
remaining IR sensitivity in the difference between the exact and
the Monte Carlo NLO expression does not have sensible consequences.
This last problem is in fact, in practice, a very minor problem.
\item
The NLO correction needs not be positive. Thus, one is forced
to accept negative weighted events.
\end{itemize}

If the hardest emission in a shower Monte Carlo is generated first, one
has the possibility of replacing the first emission with the exact
one. We now illustrate schematically how this can be done.
For ease of illustration, we assume that the Born final state
is described by the variables $v_1\ldots v_l$, and the final state
with a real emission is described by $v_1\ldots v_l, r_1, r_2, r_3$,
where the variables $r_i$ are associated with the radiated parton.
We assume that the projection is simply ${\mathbb P}\{v_1\ldots v_l, r_1, r_2, r_3\}
\to \{v_1\ldots v_l\}$.
The phase space is written as $d\Phi_{m+1}=d\Phi_v\, d\Phi_r$
where $d\Phi_v$ is the Born phase space, and $d\Phi_r$ is $\Pi dr_i$ times
a suitable Jacobian.
We now write the NLO exact formula in the following way
\begin{eqnarray}\label{eq:exactnlonf}
d\sigma \;=\; B(v) d\Phi_v + V(v) d\Phi_v
+
\left[R(v,r) d\Phi_v d\Phi_r
-C(v,r) d\Phi_v d\Phi_r {\mathbb P}
\right]
\;=
\nonumber \\
\left[V(v)
+ \left(R(v,r)
- C(v,r)\right) d\Phi_r {\mathbb P}\right]  d\Phi_v
+
B(v) d\Phi_v \left[1+
\frac{R(v,r)}{B(v)}\left(1-{\mathbb P}\right)
d\Phi_r \right]
\end{eqnarray}
Comparing eqs.~(\ref{eq:hardestapprox}) and
(\ref{eq:mcnlo}), we immediately see that the analogue of eq.~(\ref{eq:hardestapprox})
arising from eq.~(\ref{eq:exactnlonf}) is given by
\begin{eqnarray}\label{eq:nlofirstemission}
d\sigma&=&\left[V(v)
+ \left(R(v,r)
- C(v,r)\right) d\Phi_r {\mathbb P}\right]  d\Phi_v
\nonumber\\
&+&
B(v) d\Phi_v  \left[\Delta_R^{({\rm NLO})}(0)+
\Delta_R^{({\rm NLO})}(\pt)
\frac{R(v,r)}{B(v)}
 d\Phi_r \right]
\end{eqnarray}
where we have defined
\begin{equation}\label{eq:nlofirstemissionsud}
\Delta_R^{({\rm NLO})}(\pt)
=e^{-\int d\Phi_r\frac{R(v,r)}{B(v)}
\theta(\kt(v,r)-\pt)}
\end{equation}
One can implement eq.~(\ref{eq:nlofirstemission}) in an {\SMCpNLO}
implementation by generating Born events with distribution
$B(v_1\ldots v_l)$, generating the first emission according to the
second line of eq.~(\ref{eq:nlofirstemission}), and then generating
the subsequent emissions as $\pt$ vetoed shower. Furthermore,
one should associate a truncated vetoed shower from the combined
emitted parton and the closest (in $\pt$) primary parton.
The first term in eq.~(\ref{eq:nlofirstemission}) can be generated
independently, and attached to an ordinary shower, since it is formally
of higher order in $\alpha_S$. With this method, negative weighted
events could be generated, since this term is not guaranteed to be positive.
A better procedure would be the following. One defines
\begin{eqnarray}\label{eq:bbardef}
\bar{B}(v)
&=&B(v)+V(v)
\nonumber \\
&+& \int \left(R(v,r)
- C(v,r)\right) d\Phi_r
\end{eqnarray}
and then implements the hardest emission as
\begin{equation}\label{eq:nlofirstemissionp}
d\sigma=
\bar{B}(v) d\Phi_v  \left[\Delta_R^{({\rm NLO})}(0)+
\Delta_R^{({\rm NLO})}(\pt)
\frac{R(v,r)}{B(v)}
 d\Phi_r \right]\;.
\nonumber
\end{equation}
Eq.~(\ref{eq:nlofirstemissionp}) overcomes the problem of the negative
weights, in the sense that the region where
$\bar{B}$ is negative must signal the failure of perturbation
theory, since the NLO negative terms have overcome
the Born term.

The structure of the counterterm and the projection in NLO calculations
is in general more involved than in the example illustrated above.
However, one can separate the real contribution
into several term, each one of them singular in a particular collinear
region\footnote{For example, defining
$R_k=\frac{1}{\sum_i \frac{1}{S_i}}\frac{1}{S_k}$, where $S_k$ is the mass of the
pair formed by the $k^{\rm th}$
parton with the radiated parton, we have $\sum R_i=R$, and each $R_k$ is singular
only in the region where the emitted parton is collinear to the $k^{\rm th}$ parton, or soft.}.
To each term one can associate a counterterm with a projection, and
choose $v$ and $r$ variables in such a way that the projection leaves
the $v$ variables unchanged, and the phase space has the factorized form $d\Phi_v d\Phi_r$.
The discussion given above would go through unchanged, except that
in the exponent of (\ref{eq:nlofirstemissionsud}) and in (\ref{eq:nlofirstemissionp}) several
terms would appear, one for each singular region, instead of the single $R/B$ term,
in closer analogy with eqs.~(\ref{eq:multisud}) and (\ref{eq:hardestapprox}).
\subsection{Alternative formulations}
Several variants of eq.~(\ref{eq:nlofirstemissionp}) are possible. First of all,
one may replace $B$ with $\bar{B}$ in the square bracket and in the Sudakov form factor,
the difference being of NNLO order.

It is not strictly necessary to have the exact expression for $R$ in
eqs.~(\ref{eq:nlofirstemissionsud}) and
(\ref{eq:nlofirstemission}). We can choose an $\tilde{R}\le R$, and
add back the positive difference $R-\tilde{R}$ with standard Monte
Carlo methods, provided that the difference is non-singular, i.e. that
$\tilde{R}$ has the same singularity structure of $R$. For example, we
may want $\tilde{R}$ to smoothly vanish outside of the singular
regions, in order to avoid the unnecessary exponentiation of large NLO
corrections unrelated to the singular region.  This is easily obtained
by defining $\tilde{R}=R \times h(\pt)$ where $h(\pt)$ goes to one as
$\pt\to 0$ and vanishes more or less rapidly when $\pt$ is away from
zero.  There will be a left over contribution $R-\tilde{R}$, which is
positive and non-singular, and can be added separately.

\subsection{Running coupling}
In standard NLO calculations one usually uses a coupling constant evaluated
at a fixed renormalization scale, of the order
of some characteristic hard scale in the process. The SMC's use instead
a running scale for the emission vertex, of the order of $\pt$.
In order for the NLO corrected hardest emission not to spoil the all order features
of the SMC, it is convenient therefore to perform an analogous
choice in eq.~(\ref {eq:nlofirstemissionp}).
This can be easily done by taking equal to $\pt$
the argument for one power of $\as$ (i.e. the one associated with the next-to-leading order radiation)
in the expression for $R$ wherever it appears inside the square bracket of eq ~(\ref {eq:nlofirstemissionp})
(i.e. also in $\Delta_R^{(\rm NLO)}$).
By doing this one gets the correct
LL Sudakov suppression of events with a small
largest $\pt$. It is easy to see that this does not spoil the NLO accuracy of the result,
since it affects a term which is already of next-to-leading order.
\subsection{Emission from fermions}
\begin{figure}[htb]
\begin{center}
\epsfig{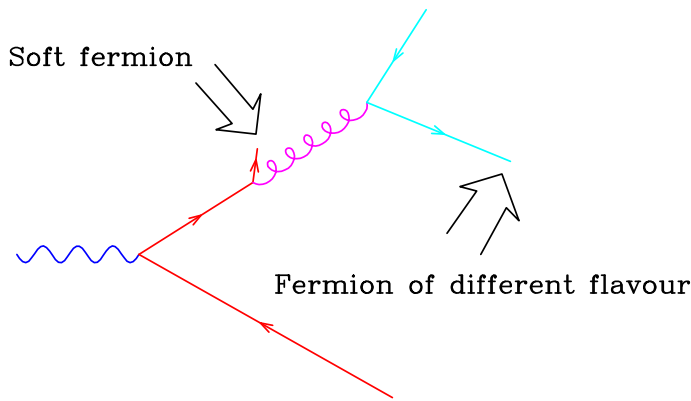}
\end{center}
\caption{\label{fig:followfermions}
If one follows the hardest line instead of the fermion line, a hardest
emission which is not present at the NLO level may be generated.}
\end{figure}

Following the hardest line in order to find the hardest emission
is only necessary for gluon emission lines. Since all emissions before the hardest
must be soft, if the soft particle emitted is not a gluon
one has power suppression of the corresponding configuration, which can therefore be
discarded. Thus, in the case of a fermion line, one can simply follow the fermion line.
In fact, by not doing so one generates configurations that are not even present in the NLO
corrections. The example of fig.~\ref{fig:followfermions} clarifies
this issue.
If one follows the hardest line instead of the fermion line one can have a hardest emission
of a fermion of different flavour, which is not present at the NLO level.
For the same reason, if one reaches a $g\to q\bar{q}$ splitting while following
a gluon line, one should not proceed any further. In fact, if the hardest emission
arises after this, according to our statement (II) the $1-z$ value of the gluon splitting is
forced to be small,
thus yielding a suppressed configuration, since the gluon splitting process is not soft-singular.
\section{Initial state showers}\label{sec:initshowers}
\subsection{Kinematics}
\begin{figure}[htb]
\begin{center}
\epsfig{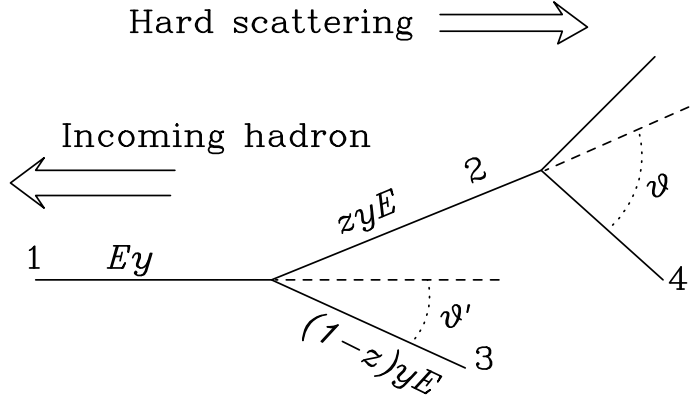}
\end{center}
\caption{\label{fig:spacelike-kin}
Kinematic variables for a spacelike splitting process.}
\end{figure}

We will now study the Monte Carlo equation for an initial state
shower. We denote with $\bar{\bS}_i(t,x)$ the backward shower from
a parton of type $i$, momentum fraction $x$ and scale variable $t$.
The kinematic is represented in fig.~\ref{fig:spacelike-kin}.
The emissions are ordered in increasing angles towards the hard scattering.
Also in the case of initial state showers the angular ordering implements
interference effects in soft gluon emission.
In fig.~\ref{fig:spacelike-kin}, for example,
in the region where $\theta\gg\theta^\prime$ and for soft (gluon) emission,
particle 4 is emitted coherently from particles 1 and 3, and since the
charge difference of 1 and 3 is the charge of 2 it appears as if it was
emitted from 2.

We consider the region where $x$ (and therefore also all $z$'s) is not small,
that is to say we ignore the small-$x$ problem now. The evolution variable
can then be taken to be
\begin{equation}\label{eq:backevvar}
 t=E^2\theta^2\;,
\end{equation}
so that
\begin{equation}
\pt=\sqrt{t} y z(1-z)\,.
\end{equation}
\subsection{Backward evolution}
In the backward evolution formalism \cite{Sjostrand:1985xi}
the probability of evolving backward
from $t,x$ to $t^\prime,x$ without emitting is given by~\cite{Marchesini:1988cf}
\begin{equation}
\Pi(t,t^\prime,x)=\frac{f(x,t^\prime)}{f(x,t)}\,\Delta(t,t^\prime)
=\frac{f(x,t^\prime)\,\Delta(t)}{f(x,t)\,\Delta(t^\prime)}\;,
\end{equation}
where $f$ is the parton density.
It is thus the same as in the timelike case, except for the
factor $f(x,t^\prime)/f(x,t)$, which accounts for the different probability
to find a parton with a fraction $x$ of the incoming momentum at the
scale $t^\prime$ instead of the scale $t$. As $t^\prime\to t_0$ the probability
$ \Pi(t,t^\prime,x)$ goes to its minimum,
and for $t^\prime\to t$ it goes to 1.
In order for the formalism to work $\Pi$ must also be a monotonic function
of $t^\prime$. It is interesting to see how this works, by taking
the derivative of $\Pi(t,t^\prime,x)$ with respect to $t^\prime$
\begin{equation} \label{eq:backevdiff}
t^\prime\frac{\partial}{\partial t^\prime}
\Pi(t,t^\prime,x)=
\frac{\Pi(t,t^\prime,x)}{f(x,t^\prime)}
\int dz\,\left[ t^\prime F(t^\prime,z) f(x,t^\prime) 
+\frac{\as}{2\pi} P(x/z) f(z,t^\prime) \frac{1}{z} \theta(z-x) \right]\;,
\end{equation}
where we have used the Altarelli-Parisi equation for $f$. Since
\begin{displaymath}
F(t^\prime,z)=\frac{\as}{2\pi t^\prime} \hat{P}(z)
\end{displaymath}
we see that the first term on the right hand side of eq.~\ref{eq:backevdiff}
is exactly what is needed to turn the regularized splitting kernel
$P(x/z)$ in the second term into the unregularized one
(thus leaving a positive definite expression in the square
bracket) provided the argument of $\as$ in the evolution equation
matches the one used in $F(t,z)$. Notice that we have also assumed that
the evolution variable
of the Altarelli Parisi equation is the same variable
used in the shower evolution. With these assumptions
we thus have
\begin{equation} \label{eq:backevdiff1}
\frac{\partial}{\partial t^\prime}\Pi(t,t^\prime,x)=
\Pi(t,t^\prime,x) \frac{1}{f(x,t^\prime)}\int_x^1 \frac{dz}{z}
F(t^\prime,x/z) f(z,t^\prime) \;.
\end{equation}
From eq.~(\ref{eq:backevdiff1}) we derive the alternative form
of the no-branching probability in backward evolution
\begin{equation} \label{eq:backsudakov}
\Pi(t,t^\prime,x)=\exp\left\{ - \int_{t^\prime}^t dt^{\prime\prime}
 \int \frac{dz}{z}\, F(t^{\prime\prime},x/z)
\frac{f(z,t^{\prime\prime})}{f(x,t^{\prime\prime})}\right\}\;,
\end{equation}
which is the form proposed in the paper where
backward evolution was first introduced \cite{Sjostrand:1985xi}.

The spacelike shower equation for backward evolution is given by
\begin{eqnarray} \label{eq:backevol}
&&
\bar{\bS}_i(t,x)=\Pi_i(t,t_0,x)
+
\\ \nonumber
&&\int_{t_0}^t dt^\prime\,dz\,dy\,
\Pi_i(t,t^\prime,x)\,
F_{ij}(t^\prime,z)\delta(x-zy) \bar{\bS}_j(t^\prime,y)\,
\frac{f_j(y,t^\prime)}{f_i(x,t^\prime)}
\bS_{l(ij)}(t^\prime (1-z)^2,y(1-z))\;.
\end{eqnarray}
The notation $\bar\bS$ stands for the spacelike shower, and $l(ij)$
stands for the other parton in the splitting $ij$
(i.e., for example $l(qq)=g$, $l(gq)=\bar{q}$, etc.).
Observe that the starting value of the ordering variable after the splitting
is exactly $t^\prime$, which is consistent with our choice of the ordering
variable for backward evolution, eq.~(\ref{eq:backevvar}).
The initial condition for the final state shower is instead consistent
with our conventions for timelike showers.

It is instructive to see how unitarity arises in eq.~(\ref{eq:backevol}).
One has
\begin{equation}\label{eq:backevoluni}
f_i(x,t)=\Delta_i(t,t_0)\,f_i(x,t_0)
+\int_{t_0}^t dt^\prime\,dz\,dy\,\Delta_i(t,t^\prime)\,
F_{ij}(t^\prime,z)\delta(x-zy)\,f_j(y,t^\prime)\;.
\end{equation}
Eq.~(\ref{eq:backevoluni}), together with the definition of the Sudakov
form factors\footnote{The factor of $1/2$ in front of $P_{gg}$
in eq.~(\ref{eq:sudform}) is appropriate for the exclusive splitting functions
that appear in the Sudakov form factors. It is the symmetry factor
associated with the identical
particles in the final state of the gluon splitting process.
For the same reason only the $P_{qq}$ function appears in $\Delta_q$.
In fact, one might replace $P_{qq}\to [P_{qq}+P_{gq}]/2$
in $\Delta_q$, and $P_{qg}\to [P_{qg}+P_{\bar{q}g}]/2$ in
$\Delta_g$, to have more symmetric expressions.}
\begin{eqnarray}
\Delta_q(t,t_0)&=&
\exp\left[-\int_{t_0}^t \frac{dt^\prime}{t^\prime} \int_0^1 dz
\frac{\as}{2\pi} P_{qq}(z)\right]\,, \nonumber \\ \label{eq:sudform}
 \Delta_g(t,t_0)
&=&\exp\left[-\int_{t_0}^t \frac{dt^\prime}{t^\prime}  \int_0^1 dz
\frac{\as}{2\pi}\left\{\frac{1}{2} P_{gg}(z) + n_f P_{qg}\right\}
  \right]\,
\end{eqnarray}
yields the regularized Altarelli-Parisi equations
\begin{equation}\label{eq:herwigAP}
\frac{d}{dt} f_i(x,t)=
\int_x^1\frac{dz}{z}\, \hat{F}_{ij}(t,z)
f_j(x/z,t)\;.
\end{equation}
\subsection{The SMC parton densities}
Eq.~(\ref{eq:herwigAP}) is a formulation of the Altarelli Parisi
equation which is appropriate for angular ordered
showers \cite{Marchesini:1988cf,Catani:1991rr}.
It differs from the traditional formulation in the definition of the
momentum variables, and also in the $z$ dependent
argument of $\as$. The corresponding parton density $f^{\rm HW}$
are equivalent to the $\overline{\rm MS}$ parton densities
$f^{\rm \overline{MS}}$
in collinear leading order.
They however differ by dominant double log terms in the threshold region.
In order to compare them, we write the expression of the
Drell-Yan cross section that they yield.
In the HERWIG scheme, the initial evolution angle for Drell-Yan is of order 1,
and the energy of order $Q$ (the square root of the lepton pair
invariant mass), so that the initial $t$ value is $Q$.
Using the HW parton densities and the Born term for the
partonic cross section, we get schematically, for the $N^{\rm th}$
moment of the cross section,
\begin{equation}\label{eq:dyherw}
\sigma_N^{\rm DY}(Q)=\hat{\sigma}_N^{\rm DY} \left[f_N^{\rm HW}(Q)\right]^2\;,
\end{equation}
where $f_N$ stands for the $N^{\rm th}$ moment of the parton density.
On the other hand, in the $\overline{\rm MS}$ scheme we have
\begin{equation}\label{eq:dymsb}
\sigma_N^{\rm DY}(Q)=\hat{\sigma}_N^{\rm DY} {f_N^{{\rm
\overline{MS}}}}^2(\mu) \exp\left\{ -\frac{4\Cf}{\pi} \int_0^1
dz\frac{z^N-1}{1-z} \int_{Q^2(1-z)^2}^{\mu^2} \frac{dq^2}{q^2} \as(q^2)
\right\}\;,
\end{equation}
which also includes double log resummation in the large $N$ limit \cite{Catani:1991rr}.
Equations (\ref{eq:dymsb}) and (\ref{eq:dyherw}) have the same $Q^2$
dependence
\begin{equation}
Q^2\frac{\partial}{\partial Q^2}\sigma_N^{\rm DY}(Q)
=\sigma_N^{\rm DY}(Q)\frac{4\Cf}{\pi} \int_0^1
dz\frac{z^N-1}{1-z}  \as(Q^2(1-z)^2)\;,
\end{equation}
which can be derived directly from eq.~(\ref{eq:dymsb})
for the $\overline{\rm MS}$ scheme,
and from eqs.~(\ref{eq:dyherw}) and (\ref{eq:herwigAP}) for the HERWIG scheme.
This suggests the following identification
\begin{equation}
f_N^{\rm HW}(t)=f_N^{\rm \overline{MS}}(\mu)
\exp\left\{ -\frac{2\Cf}{\pi} \int_0^1 dz\frac{z^N-1}{1-z}
\int_{t(1-z)^2}^{\mu^2} \frac{dq^2}{q^2} \as(q^2) \right\}\;.
\end{equation}
Thus, it is possible to define the HW parton densities in such a way
that they resum all Sudakov double logs arising in initial state radiation.
It is not possible, however, to define them to include all double logs.
For example, in the DIS
process there are double logs arising from the final state
jet, due to the fact that as $x\to 1$ the final state jet mass is forced to become
small, thus inducing further Sudakov suppression.
This leads to the well known fact that it is not possible
to absorbs soft double logs universally in the parton densities.

If the final state jet is described by an angular ordered shower,
the soft gluon structure of the final state is also correctly
described in the DIS case. However, in order to get an accurate
total rate in the threshold region, one has to correct the
parton cross section with the Sudakov term associated to the
final state jet \cite{Catani:1991rr}.

When implementing an {\SMCpNLO} program with hadrons in the initial
state, one should remember that, in order to have NLO accuracy,
NLO structure functions should be used. On the other hand, we have
seen that for consistency a HERWIG scheme should be used for
the structure functions. When discussing NLO corrections we will
have to pay attention to this problem, and we will see in the following
that it may be overcome in several ways.

\subsection{Hardest emission}
In the spacelike, backward evolution shower it is obvious that the
hardest emission is to be found along the spacelike line. An
arbitrary number of soft emissions can occur before
the hardest one. They can be implemented (as in the timelike
case) as truncated, vetoed backward showers. After the hardest emission
the backward shower will continue with vetoed emissions.

Having singled out the hardest
emission, one has an alternative formulation of the angular ordered
shower, in which the hardest emission is generated first.
It would be generated with the HERWIG parton densities, and a transverse
momentum Sudakov form factor. Thus,
it would include some NLO corrections, more precisely the threshold
corrections arising from initial state radiation.
As discussed above, it would not include other threshold corrections,
like those arising in DIS from final state jets.
One can now correct the hardest emission, using the exact matrix
elements and NLO structure functions.
This will guarantee NLO accuracy for IR and
collinear safe quantities. On the other hand, the rest of the initial
state shower can be implemented using the LL HERWIG
structure functions. This does not affect IR safe quantities, and
on the other hand guarantees consistency in the treatment of
soft radiation from the initial state line.

The equation for the angular ordered initial state shower
can be formally solved by iteration.
The hardest emission can be singled out as in eq.~(\ref{eq:largestptexp}),
yielding
\begin{equation}\label{eq:largestptback}
\bar{\bS}(t_I,x)=\Pi(t_I,t_0,x)\bI+
\sum_{l,k=0}^{\infty}
\mbox{\epsfig{file=largestptback.eps,,width=0.6\textwidth}}\;.
\end{equation}
We will also use the notation $z_i/t_i/y_i$ for $i=1,l+k+1$ to denote collectively
the $z_i/t_i/y_i$, $z/t/y$ and $\tilde{z}_i/\tilde{t}_i/\tilde{y}_i$ sequences.
The graphic expression in eq.~(\ref{eq:largestptback}) has the following
meaning.
\begin{itemize}
\item
Thick lines are Sudakov form factors; more specifically,
they are given by
\begin{equation}
 \Pi(t_I,t_1,x),\,\Pi(t_1,t_2,y_1),\ldots,
\Pi(t_{k+l+1},t_0,y_{k+l+1})\,.
\end{equation}
\item
Each $y$ value equals the $y$ value to its left divided by $z$.
Thus $y_1=x/z_1$, $y_2=y_1/z_2$, etc.
\item
The solid blobs are $\bS(y_i^2(1-z_i)^2 t_i)$.
\item
The hollow blobs stand for
\begin{equation}
F(z_i,t_i)\,\frac{f(y_i,t_i)}{f(z_i y_i,t_i)}\,
\theta(\pt - \sqrt{t_i} y_i z_i(1-z_i))\;.
\end{equation}
\item
The solid vertex blob (i.e. the largest $p_T$ vertex) is given by the
same expression as the hollow blob without the theta function.
\item
All intermediate $z_i$, $t_i$ are integrated.
\end{itemize}
As in the timelike case, eq~(\ref{eq:largestptback}) does not
have a simple Monte Carlo implementation, since no-branching factors
do not match the branching factors. We can however introduce modified
form factors, so that they match the splitting probability,
and collect the leftover. More precisely, using the expression
of eq.~(\ref{eq:backsudakov}) for $\Pi$, we rewrite it as
\begin{equation}
\Pi(t,t^\prime,x)=\Pi_V(t,t^\prime,x)
\times
\exp\left\{ - \int_{t^\prime}^t dt^{\prime\prime}
 \int \frac{dz}{z}\, F(t^{\prime\prime},z)
\frac{f(x/z,t^{\prime\prime})}{f(x,t^{\prime\prime})}
\theta(\sqrt{t^{\prime\prime}} x (1-z)-\pt)
\right\}\;,
\end{equation}
with
\begin{equation}
\Pi_V(t,t^\prime,x)=
\exp\left\{ - \int_{t^\prime}^t dt^{\prime\prime}
 \int \frac{dz}{z}\, F(t^{\prime\prime},z)
\frac{f(x/z,t^{\prime\prime})}{f(x,t^{\prime\prime})}
\theta(\pt-\sqrt{t^{\prime\prime}} x (1-z))
\right\}\;.
\end{equation}
Now $\Pi_V$ matches exactly the emission vertex. It can be easily implemented
by generating the next evolution scale with the usual form factor, supplemented
by a $\pt$ veto procedure.
The remaining factors should be collected together, yielding
\begin{equation}\label{eq:backremnant}
\Pi_{\rm rem}=
\exp\left[-\sum_{i=0}^{l+k+1}\int_{t_{i+1}}^{t_i}dt\,\frac{dz}{z}\,F(t,z)
\frac{f(y_i/z,t)}{f(y_i,t)}\theta(\sqrt{t}y_i(1-z)-\pt)\right]\,.
\end{equation}
The $\Pi_{\rm rem}$ term is easily shown to be equivalent to
\begin{equation}\label{eq:pirdef}
\Pi_{\rm rem} \approx \Pi_R(t_I,t_0,x)=
\exp\left[-\int_{t_0}^{t_I}dt\,\frac{dz}{z}\,F(t,z)
\frac{f(x/z,t)}{f(x,t)}\theta(\sqrt{t}x(1-z)-\pt)\right]\,.
\end{equation}
In fact, the theta function in eq.~(\ref{eq:backremnant}) prevents the appearance
of any large (soft or collinear) log when $t\approx \pt$. Now, if there is an $i$
for which $z_i$ is not near one, for all $j\geq i$ we have
$t_i\lesssim \pt$. In fact, at the $i^{\rm th}$ vertex we must have
\begin{equation}
t_i y_i z_i (1-z_i)\leq \pt \Longrightarrow
 t_i \lesssim \pt\;.
\end{equation}
But for all $t_j$ with $j>i$ we have $t_j<t_i$ because
of angular ordering, and so also $t_j\lesssim \pt$.
It follows that in eq.~(\ref{eq:backremnant}), for all terms in the sum
that are not suppressed by the $\theta$ function all previous $z_i$ values are
near 1, and so the corresponding $y_i$ value is near $x$. All terms in the
sum have then the same integrands, and the integration ranges can be joined together
to yield eq.~(\ref{eq:pirdef}).
\subsection{NLO Correction}
The {\SMCpNLO} correction to the hardest emission is obtained as in the timelike case,
and in fact it has the same expression given in eqs.~(\ref{eq:nlofirstemission}) and
(\ref{eq:nlofirstemissionsud}),
provided one includes in the definition of $B$, $V$, $C$ and $R$ the appropriate
structure function factor. By doing so, in the collinear regions the $R/B$ expression
entering in the Sudakov form factor (\ref{eq:nlofirstemission}) matches the expression
in eq.~(\ref{eq:pirdef}).

As in the case of the timelike shower, one can use $\pt$ as the argument of one power of
$\as$ in the expression for $R$ appearing in the square bracket of eq.~(\ref{eq:nlofirstemissionp}).
In the same instances one can also replace the NLO parton densities with the HERWIG ones, at the price of
a NNLO unknown effect, and with the advantage of a treatment of the all order
double log region which is consistent with the original SMC. Thus, it is possible
to perform the matching in a fully consistent way. Only practical studies of real
implementations will tell whether all these subtleties are of any importance at all.

\section{Comparison with other methods}\label{sec:other}
I will now discuss comparisons and relations with few popular methods
used to perform some sort of NLO improvement of the SMC.

\subsection{Matrix element corrections}
With this method one corrects the large angle emissions. Thus, large angle gluon
emission become accurate at the NLO level, but the full integral of emissions at any
angle has still only LO accuracy. A discussion of this method appropriate to SMC
with interfering gluons has been introduced in ref.~\cite{Seymour:1994df}.
One generates the shower as usual, and applies a correction factor corresponding to
the exact tree level expression divided by the SMC approximate one only if the emission
is the hardest that has taken place so far.
In order for this method to work, one has to make sure that the SMC fills all the phase
space for gluon emission, in order not to miss any region.

The method proposed in the present work can achieve the same purpose, just by
using $B$ instead of $\bar{B}$ in eq.~(\ref{eq:nlofirstemissionp}) so that
one just needs to know the matrix elements for real emission.
It has the advantage that one needs not to worry about filling the phase space for gluon emission.

\subsection{Tree level initial state with radiation}
\begin{figure}[htb]
\begin{center}
\epsfig{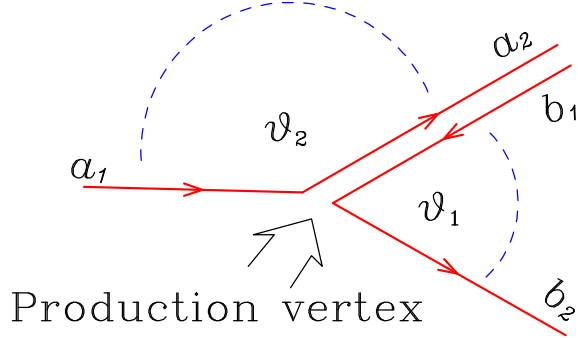}
\end{center}
\caption{\label{fig:epemginit}
Radiation from the $q\bar{q} g$ system in the planar approximation.}
\end{figure}

In the SMC generation of $e^+e^-\to hadrons$ events
one usually starts with a $q\bar{q}$ event as the hard event.
When three jet events are studied, one can start with a $q\bar{q}g$ event, requiring
a cut on some final state variable (like thrust) in order to stay away from the two-jet region.
In SMC including gluon interference, one has then to set appropriate initial showering
angle for each final state line. This poses no problems when the initial state lines
form large angles. It may cause problems if some partons form relatively small angles.
To illustrate the problem, we consider the $e^+e^-\to q\bar{q} g$ example.
We draw the colour lines for this process in the planar approximation, as shown in
fig.~\ref{fig:epemginit}.

The gluon has a colour and anticolour line associated with it.  Thus
its emission angle can be controlled either by $\theta_1$ or by
$\theta_2$. On the other hand, in the angular ordered formalism
one generates a single shower starting from a given parton, with
a given initial angle.
The HERWIG program chooses the initial condition for the evolution
of the gluon jet by picking $\theta_1$ or $\theta_2$ with equal probability.
In both cases the radiation probability of the gluon line
is given by the gluon coupling $C_A$. This works properly for the
next soft emission in the shower.
In fact, let us examine the soft emission pattern of the $q\bar{q}g$ system
in the large $N$ limit.
It is given by the incoherent sum of the emissions from the two antennas
$a_1,a_2$ and $b_1,b_2$.
The $b$ antenna only radiates inside the small angle (i.e. $\theta_1$)
cone in the $qg$
direction. The $a$ antenna fills the whole solid angle.
The colour charge of the gluon's colour lines $a_2$ and $b_1$
is $C_F$. Near the gluon, for emission angles below $\theta_1$ 
both antennas give a collinear
enhanced contribution of the same size,
and their sum correspond to replacing $C_F\to 2C_F\approx C_A$.
In HERWIG, near the gluon (i.e. for angles in the $\theta_1$ cone)
one has in fact radiation with charge $C_A$.
Near the gluon, but for emission angles larger than $\theta_1$,
only 50\% of the events radiate with $C_A$ intensity.
The net effect is radiation of the $qg$ system
with a $C_F$ intensity in the large angle region. This is what we expect:
the $a$ antenna radiates with intensity $C_F$, and the $b$ antenna does not radiate
for angles larger than $\theta_1$.
This clever scheme, however, does not work for further emissions.
In fact, lack of soft radiation around the $qg$ system should
be suppressed by a Sudakov form factor. In the HERWIG scheme
we instead have just a 50\% suppression.
In fact, in 50\% of the events the $qg$ system does not radiate
for angles greater than $\theta_1$.
This means that for pseudorapidity below the $\theta_1$ angular region
the $qg$ system would be scarcely populated
by soft radiation. It is not clear whether this fact may cause
observable problems. At the end of the shower the coloured parton
associated with the gluon colour line will have a colour partner
originating from soft radiation from the $a$ antenna.
According to the argument given above this partner
may not be close enough in rapidity to form a low mass
colour cluster.

The method proposed in this paper suggests instead to assign
an initial showering angle of $\theta_1$ to both the quark
and the gluon.
One then adds a truncated
shower from $\theta_2$ to $\theta_1$ associated with the $qg$
system. It is clear that this procedure gives the right
amount of soft radiation colour connected to the outgoing gluon
line.
\subsection{The CKKW formalism}
The CKKW formalism \cite{Catani:2001cc} is a method to merge
perturbative angular ordered showers with exact matrix elements.
The method requires a cutoff on the cluster mass, which is used
to separate the application of exact and approximate
matrix elements.

There are several analogies among concepts that have arisen in the
present work and in the CKKW paper.  In particular, the need to
introduce vetoed showers arises also there.

Here I would like to focus on one concept that has no parallel
in the CKKW work: the concept of a truncated
shower. In fact, CKKW avoid the need of a
truncated shower using a trick\footnote{See Sec. 3.2 of \cite{Catani:2001cc}.}
that I will illustrate here referring to the $e^+e^-$
example of fig.~\ref{fig:epemginit}.

In the CKKW method
one limits the gluon radiation to $\theta_1$, but
lets the quark radiate at all angles.
In the truncated shower scheme
one requires that both the gluon and the quark radiate starting
from the angle $\theta_1$, and then adds a truncated shower
of the $qg$ system for larger angles. Since the colour charge of
the $qg$ system is equal to the charge of the quark, the radiation
of the truncated shower combines with the radiation of the quark
to yield quark radiation in the whole solid angle, which
is the CKKW method.
This also works for gluon splitting:  one lets the harder gluon
radiate at all angles. The $g\to q\bar{q}$ has to be treated
differently. Here one should let both quark and antiquark radiate
at large angle, so that one reconstructs the $C_A$ charge of the
truncated shower by summing up two $C_F$'s.

As far as particle flow and multiplicities, the CKKW method is
equivalent to the truncated shower method. It is however different
in the colour connection that it assigns to final state partons.
The colour connection of a soft emission in the truncated shower is
the same as if the emission came from the parent parton.  This is what
it should be, since it is coherent radiation.  In the CKKW method
colour connections are not respected, and this may cause problems
with large mass colour connected parton clusters.

\section{Conclusions}\label{sec:conclusions}
In the present paper I presented a method for the inclusion
of NLO corrections in Monte Carlo programs with angular ordering,
that avoid the problem of negative weighted event. 
The method may be applied also for the less ambitious goal of
correcting the large angle emissions in a SMC. Also
in this contexts it has some advantages.
The implementation of this method requires minor modifications
to the showering scheme in SMC programs. One modification is
the inclusion of $p_T$ vetoing in the shower.
This is common practice in SMC, and should not present
particular problems. Truncated showers instead have
never appeared in SMC implementations. They play a central
role in the present work, since it does not seem possible
to implement the soft radiation of a collinear
bunch of partons without truncated showers. It is therefore
desirable that in the context of rewriting or developing
new Monte Carlo programs will provide facilities to
implement truncated showers.
Besides opening the possibility of {\SMCpNLO} implementations
for several processes, this would also allow to improve
matrix element corrections schemes with a limited effort.

The method suggested here also hints at the possibility of SMC+NNLO
implementations. One should isolate the two hardest emission
in a SMC event, consider again $\pt$ vetoed showers at the
smaller $\pt$ of the two hardest emissions, and add truncated
shower associated with two or three partons arising from the
hardest splittings. Although it may be premature to discuss this
problem in details now, it may be convenient to consider
the possibility of implementing truncated shower from more
than two lines when building SMC programs.

The method discussed in the present paper has been developed
in the context of angular ordered showers, in particular as
implemented in the HERWIG SMC. The aim of this work was to provide
an approach along the lines of
ref.~\cite{Frixione:2002ik,Frixione:2003ei,Frixione:2003ep},
improved over some substantial aspects.
Different approaches to the problem of NLO improvement
of SMC programs have been presented in refs.~\cite{Collins:2001fm,Collins:2000qd,Kramer:2003jk,Soper:2003ya}.

\section{Acknowledgments.}
I would like to thank S.~Frixione, S.~Gieseke for useful conversation,
and in particular B.~Webber for patiently answering many
questions on SMC physics.
\bibliography{paper.bib}
\end{document}